\documentstyle[prd,aps,preprint]{revtex}
\def\be{\begin{equation}}
\def\ee{\end{equation}}
\def\bea{\begin{eqnarray}}
\def\eea{\end{eqnarray}}
\def\ba{\begin{array}}
\def\ea{\end{array}}
\def\ben{\begin{enumerate}}
\def\een{\end{enumerate}}

\def\lll{\label}
 
\begin{document}
\newcommand{\half}{{\textstyle\frac{1}{2}}}
\newcommand{\eqn}[1]{(\ref{#1})}

\def\a{\alpha}
\def\b{\beta}
\def\g{\gamma}\def\G{\Gamma}
\def\d{\delta}\def\D{\Delta}
\def\ep{\epsilon}
\def\et{\eta}
\def\z{\zeta}
\def\t{\theta}\def\T{\Theta}
\def\l{\lambda}\def\L{\Lambda}
\def\m{\mu}
\def\f{\phi}\def\F{\Phi}
\def\n{\nu}
\def\p{\psi}\def\P{\Psi}
\def\r{\rho}
\def\s{\sigma}\def\S{\Sigma}
\def\ta{\tau}
\def\x{\chi}
\def\o{\omega}\def\O{\Omega}
\def\k{\kappa}
\def\pa {\partial}
\def\ov{\over}
\def\br{\nonumber\\}
\def\ud{\underline}

%
\title{(NS5, D5, D3) bound state, OD3, OD5 limits and $SL(2, Z)$ duality} 
\author{Indranil Mitra\footnote{E-mail: indranil@theory.saha.ernet.in} and
Shibaji Roy\footnote{E-mail: roy@theory.saha.ernet.in}}
\address
{Theory Division, Saha Institute of Nuclear Physics,\\
1/AF Bidhannagar, Calcutta-700 064, India} 
\maketitle
\begin{abstract}
We generalize the non-threshold bound state in type IIB supergravity of
the form (NS5, D5, D3) constructed by the present authors 
(in hep-th/0011236) to non-zero asymptotic value of the axion $(\chi_0$). 
We identify the decoupling limits corresponding to both the open D3-brane 
theory and open D5-brane theory for this supergravity solution as expected. 
However, we do not find any non-commutative Yang-Mills theory (NCYM) limit 
for this solution in the presence of NS5 branes.  We then study the 
$SL(2, Z)$ duality symmetry of type 
IIB theory for both OD3-limit and OD5-limit. We find that for OD3 theory, 
a generic $SL(2, Z)$ duality always gives another OD3-theory irrespective 
of the value of $\chi_0$ being rational or not. This indicates that OD3-theory 
is self-dual. But, under a special set of $SL(2, Z)$ transformations for 
which $\chi_0$ is rational OD3-theory goes over to a 5+1 dimensional NCYM 
theory and these two theories in this case are related to each other by 
strong-weak duality symmetry. On the other hand, for OD5-theory, a generic 
$SL(2, Z)$ duality gives another OD5-theory if $\chi_0$ is irrational, but 
when $\chi_0$ is rational it gives the little string theory limit indicating 
that OD5-theory is S-dual to the type IIB little string theory. 
\end{abstract} 

\newpage
\section{Introduction}

In a previous paper \cite{mitro}, we have constructed various non-threshold 
bound state solutions of both type IIB and type IIA supergravities of the
type (NS5, D$p$) (with $0\leq p \leq 5$) and (NS5, D$(p+2)$, D$p$) (with
$0\leq p \leq 3$) by applying a series of T- and S-dualities to the known
$(q,\,p)$ 5-brane solution of type IIB supergravity\footnote{Some of these
solutions are also considered in \cite{cgnn} from a different approach.}. 
One of the motivations for 
constructing such solutions is to look at the world-volume theories of 
NS5-branes in the presence of various D-branes (or various RR electric gauge
fields). In \cite{gopms}, it was argued that the world-volume 
theory of NS5-branes
in the presence of a near critical RR $(p+1)$-form electric gauge field 
gives a non-gravitational and non-local theory called an open D$p$-brane
theory in a special low energy limit (decoupling limit) known as the
OD$p$-limit. The (NS5, D$p$) brane supergravity solution in this decoupling 
limit describes the supergravity dual of OD$p$-theories. These theories are
analogous to world-volume theories of D$p$-branes in the presence of near
critical electric fields (NCOS theory) \cite{sst,gmms} and world-volume 
theory of M5-brane
in near critical electric 3-form gauge field (OM theory) \cite{gopms,bbss} 
and contains
fluctuating light open D$p$-branes in the world-volume of NS5-branes decoupled
from gravity.

Since starting from these (NS5, D$p$) bound state solutions it is possible to
construct various other bound states containing NS5-branes and several 
different D-branes by applying a series of T- and S-dualities, it is natural
to ask what kind of theories do they correspond to in the decoupling limit.
Some of the cases have been studied in \cite{ao,aos}. In this paper 
we consider a specific
case, namely, the (NS5, D5, D3) non-threshold bound state solution of type
IIB supergravity. This is an $SL(2,Z)$ invariant bound state of type IIB
theory and the supergravity solution for this state has been constructed in
\cite{mitro} for zero asymptotic value of the axion. We rewrite this
solution for the non-zero asymptotic value of the axion ($\chi_0$). This
solution can also be regarded as NS5-branes in the presence of a 6-form
and a 4-form RR electric gauge fields. There are $m$ NS5-branes, 
$n$ D5-branes and $p$ D3-branes per $(2\pi)^2 \alpha'$ of two codimensional
area of NS5 (or D5)-branes in this bound state and preserves half of the
space-time supersymmetries of string theory. We then identify both the open
D3-brane limit and the open D5-brane limit for this supergravity solution.
For the former, the 4-form approaches the critical value, whereas, for the
latter the 6-form gauge field approaches the critical value. In this paper 
we only concentrate on the supergravity solution in the decoupling limit
and also instead of writing the RR 6-form gauge field we write its Poincare 
dual. The existence of OD3- and OD5-limits\footnote{The OD3 and OD5 theory
described here are different from the usual OD3 and OD5 theory discussed
in \cite{gopms}.} for this bound state solution
may not be surprising, however, to our surprise, we do not find any NCYM-limit
in this $SL(2, Z)$ invariant solution. Since, (NS5, D3) state goes over to
(D5, D3) state under S-duality, it was argued in \cite{gopms}, that the strong 
coupling limit of OD3-theory is the (5+1)-dimensional NCYM. So, it is 
surprising that no NCYM-limit exists for (NS5, D5, D3) state. However, it 
can be easily checked that when $m = 0$ (i.e. when NS5-branes are absent),
the OD3-limit reduces exactly to the NCYM-limit \cite{aos}.

We then study the $SL(2, Z)$ transformation\footnote{$SL(2, Z)$ 
transformation on the various decoupling limits of
(F, D1, D3) bound state has been studied in \cite{lurs1,lurs2,rsh,co}.} on 
both the 
OD3-limit and the OD5-limit.
We find that under a generic $SL(2, Z)$ transformation OD3-limit always 
gives another
OD3-limit irrespective of whether $\chi_0$ is rational or not. Since
even for rational $\chi_0$, we get another OD3-limit, we conclude that 
OD3-theory is self-dual. In other words, strongly coupled OD3-theory is related
to the weakly coupled OD3-theory with different set of parameters related
by S-duality. This is in accord with recent observation made in 
\cite{lu,lars}, where
it was emphasized that since 5+1 dimensional NCYM is non-renormalizable, so
it can not be obtained from OD3-theory by S-duality. Thus OD3-theory must be
self-dual and is the UV completion of the 5+1 dimensional NCYM. But, we find
that under a special circumstance $SL(2, Z)$ duality on OD3-limit can give 
rise to NCYM-limit. In this case $\chi_0$ becomes rational and therefore, 
the OD3-theory becomes related to the NCYM-theory by the strong-weak duality. 
Actually,
what happens here is that under this special set of $SL(2, Z)$ transformations
the transformed charge of NS5-brane vanishes. Therefore, the OD3-limit
in the transformed solution reduces to NCYM-limit as we mentioned earlier.
However, for
OD5-limit, we find that when $\chi_0$ is irrational a generic $SL(2, Z)$ 
transformation
gives another OD5-limit with different parameters. But when $\chi_0$ is 
rational OD5-limit gives us precisely the little string theory 
\cite{brs,ns,dvv,lms,abks,imsy} limit. Thus
we conclude that under the S-duality of type IIB theory OD5-theory goes over
to little string theory. 
  
This paper is organized as follows. In section 2, we give the (NS5, D5, D3)
supergravity solution for non-zero asymptotic value of the axion. In section
3, we discuss the OD3- and OD5-limits for this solution. The $SL(2, Z)$
transformation is discussed in section 4. Finally, in secton 5, we present our
conclusion.
 
\section{(NS5, D5, D3) solution for non-zero $\chi_0$}

The non-threshold bound state solution of the type (NS5, D5, D3) of type
IIB supergravity was constructed in \cite{mitro} and has the form:
\bea
ds^2 &=& 
H^{1/2} H''^{1/2}\left[H^{-1}\left(-dx_0^2 + dx_1^2 + \cdots + dx_3^2\right)
+ H'^{-1}\left(dx_4^2 + dx_5^2\right) + dr^2 + r^2 d\Omega_3^2\right]\br
e^{\phi_b} &=& g_s H'^{-1/2} H''\br
\chi &=& \frac{\tan\psi}{g_s}(H''^{-1}-1)\br
B^{(b)} &=& 2m\alpha'
\sin^2\theta \cos\phi_1 d\theta \wedge d\phi_2
+\tan\varphi\sin\psi H'^{-1}
dx^4 \wedge dx^5\br
A^{(2)} &=& 2n\alpha'
\sin^2\theta \cos\phi_1 d\theta \wedge d\phi_2
-\frac{\cos\psi}{g_s}\tan\varphi H'^{-1}
dx^4 \wedge dx^5\br
A^{(4)} &=& -p\alpha' H'^{-1}\sin^2\theta \cos\phi_1
dx^4 \wedge dx^5 \wedge d\theta \wedge d\phi_2\br
&&\qquad -\frac{\sin\varphi}{g_s} H^{-1} dx^0 \wedge dx^1 \wedge dx^2 \wedge
dx^3
\lll{1}
\eea
In the above $r=\sqrt{x_6^2+x_7^2+x_8^2+x_9^2}$ and $d\Omega_3^2 = d\theta^2
+ \sin^2\theta d\phi_1^2 + \sin^2\theta \sin^2\phi_1 d\phi_2^2$ is the line
element of the unit 3-sphere transverse to the 5-branes. $g_s = e^{\phi_{b0}}$
is the string coupling constant, $B^{(b)}$ and $A^{(2)}$ denote the NSNS and
RR two-form potentials. $\chi$ is the RR scalar and $A^{(4)}$ is the RR
4-form gauge field whose field strength is self-dual. The harmonic functions
$H$, $H'$ and $H''$ are given as, 
\bea
H &=& 1 + \frac{Q_5}{r^2}\br
H' &=& 1 + \frac{\cos^2\varphi Q_5}{r^2}\br
H'' &=& 1 + \frac{\cos^2\varphi \cos^2\psi Q_5}{r^2}
\lll{2}
\eea
where the angles $\cos\varphi$, $\cos\psi$ and the charge $Q_5$ are defined as
\bea
\cos\varphi &=& \frac{(m^2 + n^2 g_s^2)^{1/2}}{[m^2 + (p^2 + n^2)
g_s^2]^{1/2}}\br
\cos\psi &=& \frac{m}{(m^2 + n^2 g_s^2)^{1/2}}\br
Q_5 &=& [m^2 + (p^2 +n^2)g_s^2]^{1/2} \alpha'
\lll{3}
\eea
Here $m$ is the number of NS5-branes $n$ is the number of D5-branes and $p$
is the number of D3-branes per $(2\pi)^2\alpha'$ of two codimensional area
of 5-branes.

Note here that since the harmonic functions in \eqn{2} approaches unity 
asymptotically, so the string metric in \eqn{1} becomes Minkowskian in this
limit. Also, $e^{\phi_b} \rightarrow e^{\phi_{b0}}$ and $\chi \rightarrow 0$
asymptotically. In order to obtain this solution for non-zero asymptotic value
of the axion $(\chi_0)$, we make an $SL(2, R)$ transformation by the matrix
\be
\Lambda = \left(\begin{array}{cc} 1 & \chi_0\\ 0 & 1\end{array}\right)
\lll{4}
\ee
The solution then takes the form:
\bea
ds^2 &=& 
H^{1/2} H''^{1/2}\left[H^{-1}\left(-dx_0^2 + dx_1^2 + \cdots + dx_3^2\right)
+ H'^{-1}\left(dx_4^2 + dx_5^2\right) + dr^2 + r^2 d\Omega_3^2\right]\br
e^{\phi_b} &=& g_s H'^{-1/2} H''\br
\chi &=& \frac{\tan\psi}{g_s}(H''^{-1}-1) + \chi_0\br
B^{(b)} &=& 2m\alpha'
\sin^2\theta \cos\phi_1 d\theta \wedge d\phi_2
+\tan\varphi\sin\psi H'^{-1}
dx^4 \wedge dx^5\br
A^{(2)} &=& 2n\alpha'
\sin^2\theta \cos\phi_1 d\theta \wedge d\phi_2
-\left(\frac{\cos\psi}{g_s} + \chi_0\sin\psi\right)\tan\varphi H'^{-1}
dx^4 \wedge dx^5
\lll{5}
\eea
and $A^{(4)}$ retains its form as given in \eqn{1} since it is $SL(2, R)$
invariant. The harmonic functions retain their form as given in \eqn{2}, but
now the angles and the charge $Q_5$ are given as,
\bea
\cos\varphi &=& \frac{[m^2 + (n + \chi_0 m)^2 g_s^2]^{1/2}}{[m^2 + (p^2 + 
(n + \chi_0 m)^2)
g_s^2]^{1/2}}\br
\cos\psi &=& \frac{m}{[m^2 + (n + \chi_0 m)^2 g_s^2]^{1/2}}\br
Q_5 &=& [m^2 + (p^2 + (n + \chi_0 m)^2)g_s^2]^{1/2} \alpha'
\lll{6}
\eea
Note here that in terms of angles $Q_5$ in \eqn{6} can be written as
$Q_5 = (m \alpha')/(\cos\varphi\cos\psi)$. Therefore, the harmonic functions
in eq.\eqn{2} take the forms:
\bea
H &=& 1 + \frac{m\alpha'}{\cos\varphi \cos\psi r^2}\br
H' &=& 1 + \frac{m\alpha' \cos\varphi}{\cos\psi r^2}\br
H'' &=& 1 + \frac{m\alpha' \cos\varphi \cos\psi}{r^2}
\lll{har}
\eea
We will use these forms later.

From eq.\eqn{6} we deduce the following quantization conditions:
\bea
\frac{n}{m} &=& \frac{\tan\psi}{g_s} - \chi_0\br
\frac{p}{m} &=& \frac{\tan\varphi}{g_s \cos\psi}
\lll{7}
\eea
Note here that under a general $SL(2, Z)$ transformation by the matrix
\be
\Lambda = \left(\begin{array}{cc} a & b\\ c & d \end{array}\right)
\lll{8}
\ee
where $a,\,b,\,c,\,d$ are integers
with $ad - bc = 1$, $Q_5$, $\cos\varphi$, $H$ and $H'$ are invariant. But
$\cos\psi$ and $H''$ change as 
\bea
\cos{\hat {\psi}} &=& \frac{\left[\frac{c}{g_s}\tan\psi - (c\chi_0 + d)\right]}
{\left[(c\chi_0 + d)^2 + \frac{c^2}{g_s^2}\right]^{1/2}} \cos\psi\br
\hat{H}'' &=& 1 + \frac{\cos^2\varphi \cos^2\hat{\psi} Q_5}{r^2}
\lll{9}
\eea
We will consider the various decoupling limits for this solution in the next
section.

\section{OD3 and OD5 limits}

\underline{{\it (a) OD3 limit:}}
The open D3-brane theory appears as a decoupling limit on the world-volume
of NS5-branes in the presence of a near critical RR 4-form electric gauge
field and was obtained in \cite{gopms}. The corresponding supergravity dual 
\cite{mitro,aor} is
given as the decoupling limit of (NS5, D3) bound state solution of type IIB
theory. The OD3-limit is given as the following:
\be
\cos\varphi = \epsilon \to 0
\lll{11}
\ee
keeping the following quantities fixed,
\be
\alpha'_{\rm eff} = \frac{\alpha'}{\epsilon}, \qquad u = \frac{r}{\epsilon
\alpha'_{\rm eff}}, \qquad G_{o(3)}^2 = g_s
\lll{12}
\ee
Note here that if we set $\cos\psi = 1$ and $\chi_0 = 0$, then the solution
given in \eqn{5} reduces to (NS5, D3) solution \cite{mitro}. However, when 
D5-branes are
also present OD3-limit is again the same as in \eqn{11} and \eqn{12}, but
in addition we set
\be
\cos\psi = l\,\, ({\rm finite}) \qquad {\rm and} \qquad \chi_0 \ne 0
\lll{13}
\ee
In the above $\epsilon$ is a dimensionless parameter, 
$(\alpha'_{\rm eff})^{3/2}$
corresponds to the finite inverse tension of open D3-brane and $G_{o(3)}^2$
is the coupling constant. In this limit the harmonic functions in \eqn{har}
reduce to
\bea
H &=& \frac{1}{a^2 \epsilon^2 u^2}\br
H' &=& \frac{h'}{a^2 u^2}\br
H'' &=& \frac{h''}{\tilde{a}^2 u^2}
\lll{14}
\eea
where $h' = 1 + a^2 u^2$ and $h'' = 1 + \tilde{a}^2 u^2$, with $a^2 = l 
\alpha'_{\rm eff}/m$ and $\tilde{a}^2 = \alpha'_{\rm eff}/(lm)$. So, the
metric in \eqn{5} takes the following form,
\be
ds^2 = \alpha' h''^{1/2}\left[-d\tilde{x}_0^2 + \sum_{i=1}^3 d\tilde{x}_i^2
+ h'^{-1}\sum_{j=4}^5 d\tilde{x}_j^2 + \frac{m}{u^2}\left(du^2 + u^2 
d\Omega_3^2\right)\right]
\lll{15}
\ee
The finite coordinates in \eqn{15} are defined as
\bea
\tilde{x}_{0,1,2,3} &=& \sqrt{\frac{l}{\alpha'_{\rm eff}}} x_{0,1,2,3}\br
\tilde{x}_{4,5} &=& \frac{\sqrt{l\alpha'_{\rm eff}}}{\alpha'} x_{4,5}
\lll{16}
\eea
So, these are precisely the OD3-limit discussed in \cite{gopms}. In this limit
the dilaton and other gauge fields take the form:
\bea
e^{\phi_b} &=& G_{o(3)}^2 \frac{h''}{h'^{1/2}}\frac{l}{\tilde{a}u}\br
\chi &=& - \frac{\sqrt{1-l^2}}{l} \frac{1}{G_{o(3)}^2}\frac{1}{h''} +
\chi_0\br
B_{\theta\phi_2}^{(b)} &=& 2m\alpha'\sin^2\theta \cos\phi_1, \qquad 
B_{45}^{(b)} = \alpha' \frac{\sqrt{1-l^2}}{l} \frac{a^2 u^2}{h'}\br
A^{(2)}_{\theta\phi_2} &=& 2n\alpha'\sin^2\theta \cos\phi_1, \qquad
A_{45}^{(2)} = -\alpha'\left(\frac{1}{G_{o(3)}^2} + \frac{\chi_0 \sqrt{1-
l^2}}{l}\right)\frac{a^2 u^2}{h'}\br
A^{(4)}_{45\theta\phi_2} &=& - \frac{\alpha'^3}{\alpha'_{\rm eff}} p
\frac{a^2 u^2}{lh'} \sin^2\theta \cos\phi_1, \qquad A_{0123}^{(4)} = 
-\frac{\alpha'^2}{G_{o(3)}^2}\frac{a^2 u^2}{l^2}
\lll{17}
\eea
Note that for $\frac{\tilde{a}u}{l} \ll 1$ (which implies that both $au \ll 1$
and $\tilde{a}u \ll 1$) i.e. in the IR region the supergravity description
is valid (in this case the curvature $\alpha' {\cal R} \sim 1/m$ remains small
for large enough $m$, the number of NS5-branes) if $\frac{\tilde{a}u}{l} \gg
G_{o(3)}^2$. In this case $G_{o(3)}^2 \ll 1$. However, this condition is
not satisfied in the extreme IR region, where $e^{\phi_b}$ becomes large. In
that case we have to go to the S-dual frame and we will describe this in the
next section. In any case, in the IR region the OD3-theory flows to (5+1)
dimensional SYM theory.

For $au \gg 1$, i.e. in the UV region, the string coupling $e^{\phi_b} =
G_{o(3)}^2 = {\rm fixed}$. So, when $G_{o(3)}^2 \ll 1$, we have valid 
supergravity description and the metric in \eqn{15} reduces to that of 
ordinary D3-branes smeared in 4, 5 directions. However, for $G_{o(3)}^2 \gg
1$, we have to go to the S-dual frame and will be discussed in the next 
section.

We would like to emphasize that the OD3 theory we are discussing here whose
supergravity dual is given in \eqn{15} and \eqn{17} are different from the
usual OD3 theory in that the field configurations in this case are different.
This is also manifested in the extra parameter defined in \eqn{13}. Note
that when the parameter $l = 1$ , it implies $n=0$ and $\chi_0 = 0$ and 
therefore the configuration \eqn{5} reduces to (NS5, D3) solution. In the
decoupling limit \eqn{11} and \eqn{12}, it will give the supergravity dual
of ordinary OD3 theory \cite{gopms,mitro,aor}. In this case we note that
$B_{45}^{(b)}$ in \eqn{17} vanishes. On the other hand if the parameter $l=0$,
it implies that $m=0$ i.e. there are no NS5 branes in the supergravity
configuration. In this case as we will explain later, the OD3 limit reduces
to NCYM limit. When $l$ is in between 0 and 1, we get a new OD3 theory
which is noncommutative as $B_{45}^{(b)} \neq 0$ and therefore the coordinates
$x^4$ and $x^5$ are noncommutative. It is clear from the expression of
$B_{45}^{(b)}$ in \eqn{17} that noncommutativity is related with the parameter
$l$ as $\sqrt{1-l^2}/l$. This gives a physical interpretation of the
parameter $l$ in this new OD3 theory.
\medskip

\underline{{\it (b) OD5 limit:}} As in the previous case open D5-brane theory
arises as a decoupling limit on NS5-branes when the RR 6-form electric gauge
field approaches a critical value. The dual supergravity solution is obtained
from (NS5, D5) bound state \cite{mitro} in the decoupling limit. In the 
present case of
(NS5, D5, D3) solution the OD5-limit is given in the following:
\be
\cos\psi = \epsilon \to 0
\lll{18}
\ee
keeping the following quantities fixed,
\be
\alpha'_{\rm eff} = \frac{\alpha'}{\epsilon}, \qquad u = \frac{r}{\epsilon
\alpha'_{\rm eff}}, \qquad G_{o(5)}^2 = \epsilon g_s
\lll{19}
\ee
and also we have
\be
\cos\varphi = \tilde{l} = {\rm finite}, \qquad {\rm and} \qquad \chi_0 \ne 0
\lll{20}
\ee
Under this decoupling limit the harmonic functions take the forms:
\bea
H &=& \frac{1}{a^2 \epsilon^2 u^2}\br
H' &=& \frac{1}{\tilde{a}^2 \epsilon^2 u^2}\br
H'' &=& \frac{h''}{\tilde{a}^2 u^2}
\lll{21}
\eea
where $h'' = 1 + \tilde{a}^2 u^2$, with $\tilde{a}^2 = \alpha'_{\rm eff}/
(m\tilde{l})$,
$a^2 = \tilde{l}\alpha'_{\rm eff}/m$.

The metric in \eqn{5} is now given by
\be
ds^2 = \alpha' h''^{1/2}\left[-d\tilde{x}_0^2 + \sum_{i=1}^5 d\tilde{x}_i^2
 + \frac{m}{u^2}\left(du^2 + u^2 
d\Omega_3^2\right)\right]
\lll{22}
\ee
where the finite coordinates are defined as
\be
\tilde{x}_{0,1,2,3} = \sqrt{\frac{\tilde{l}}{\alpha'_{\rm eff}}} x_{0,1,2,3},
\qquad \tilde{x}_{4,5} = \frac{1}{\sqrt{\tilde{l}\alpha'_{\rm eff}}} x_{4,5}
\lll{23}
\ee
The dilaton, axion and other gauge fields in the decoupling limit are given
as,
\bea
e^{\phi_b} &=& G_{o(5)}^2 \frac{h''}{\tilde{a}u}\br
\chi &=& - \frac{1}{G_{o(5)}^2}\frac{1}{h''} +
\chi_0\br
B_{\theta\phi_2}^{(b)} &=& 2m\alpha'\sin^2\theta \cos\phi_1, \qquad 
B_{45}^{(b)} = \frac{\alpha'^2}{m} \frac{\sqrt{1-\tilde{l}^2}}
{\tilde{l}} u^2\br
A^{(2)}_{\theta\phi_2} &=& 2n\alpha'\sin^2\theta \cos\phi_1, \qquad
A_{45}^{(2)} = \frac{\alpha'^2}{m}\chi_0\frac{\sqrt{1-
\tilde{l}^2}}{\tilde{l}}u^2\br
A^{(4)}_{45\theta\phi_2} &=& - \frac{p}{m}\alpha'^3 
u^2 \sin^2\theta \cos\phi_1, \qquad A_{0123}^{(4)} = 
-\alpha'^3\frac{1}{m}\frac{\sqrt{1-\tilde{l}^2}}{\tilde{l}}u^2
\lll{24}
\eea
This is precisely the OD5-limit discussed in \cite{gopms}. However, 
instead of the
RR 6-form electric gauge field, we have given here its Poincare dual. One can
indeed check that the corresponding 6-form approaches the critical value given
there in this limit.

It is clear from above that in the IR ($\tilde{a} u \ll 1$) the supergravity 
solution is valid if $\tilde{a}u \gg G_{o(5)}^2$. In that case, $G_{o(5)}^2
\ll 1$. But in the extreme IR, this relation is not satisfied and we need 
to go to the S-dual frame. On the other hand, in the UV, the solution is
valid if $\tilde{a}u \ll G_{o(5)}^{-2}$. This again is not satisfied in the
extreme UV region and we need to go to the S-dual frame which we will discuss
in the next section.

As in the case of OD3 theory, the OD5 theory whose supergravity dual is given
by \eqn{22} and \eqn{24} is also different from the usual OD5 theory 
\cite{gopms,mitro,aor}. The field contents for this new OD5 theory is 
different. In this case there is an additional parameter $\tilde{l}$ defined
in \eqn{20}. When $\tilde{l} = 1$, this new OD5 theory reduces to ordinary
OD5 theory. However, when $\tilde{l} = 0$, the configuration \eqn{5} reduces
to that of D3 brane with two additional isometries in $x^4$ and $x^5$
directions. Under the decoupling limit \eqn{18} and \eqn{19} the supergravity
description becomes invalid as the dilaton blows up. So, $\tilde{l} \neq 0$
and when it lies in between 0 and 1, we get a noncommutative OD5 theory
since $B_{45}^{(b)}$ does not vanish as can be seen from \eqn{24}. Again
the coordinates $x^4$ and $x^5$ are noncommutative. The noncommutativity
parameter in this case is proportional to 
$\alpha'\sqrt{1-\tilde{l}^2}/\tilde{l}$ i.e. it is not fixed but scales as
$\alpha'$. This should be contrasted with OD3 theory where the noncommutativity
parameter is fixed as mentioned earlier.

Thus we have obtained both the OD3-limit and OD5-limit for the (NS5, D5, D3)
supergravity solution. This as we mentioned in the introduction is quite
expected. However, contrary to our expectation, we do not find any NCYM
limit for this solution. We just like to point out that the OD3-limit 
discussed in this section takes the form of NCYM limit when we set `$m$'
the number of NS5-branes exactly to zero. Since $m = 0$, $\chi_0 = 0$,
implies $\cos\psi = 0$, so the harmonic function $H'' = 1$. Also, note
that since $m/\cos\psi = n g_s$, so, the other two harmonic functions in
the limit \eqn{11} and \eqn{12} reduce to
\bea
H &=& 1 + \frac{ng_s\alpha'}{\cos\varphi r^2} \to \frac{\tilde{b}^2}
{a^2 u^2 \alpha'^2}\br
H' &=& 1 + \frac{ng_s\alpha'\cos\varphi}{r^2} = \frac{h}{a^2 u^2}
\lll{25}
\eea
where we have defined $h = 1 + a^2 u^2$ with $a^2 = \tilde{b}/(ng_s)$. Note
that here we are interpreting $\alpha'_{\rm eff} = \tilde{b}$ as the
non-commutativity parameter and the Yang-Mills coupling $g_{NCYM}^2 =
(2\pi)^3 g_s =$ fixed. With these forms of the harmonic functions the metric
and the dilaton take precisely the same form as the dual of NCYM theory 
\cite{aos} :
\be
ds^2 = \alpha'\left[\frac{u}{R}\left(\left(-d\tilde{x}_0^2 + 
\sum_{i=1}^3 d\tilde{x}_i^2\right)
+ h^{-1}\sum_{j=4}^5 d\tilde{x}_j^2\right) + \frac{R}{u}\left(du^2 
+ u^2 
d\Omega_3^2\right)\right]
\lll{26}
\ee
where $R = \tilde{b}/a$ and the fixed coordinates are
\be
\tilde{x}_{0,1,2,3} = x_{0,1,2,3}; \qquad \tilde{x}_{4,5} = \frac{\tilde{b}}{
\alpha'} x_{4,5}
\lll{27}
\ee
and
\be
e^{\phi_b} = g_s \left(\frac{u}{R}\right)\frac{\tilde{b}}{h^{1/2}}
\lll{28}
\ee
Since NCYM limit is nothing but OD3-limit of (NS5, D5, D3) solution with
$m = 0$, so it is not clear whether there exists an NCYM limit for this
solution (for $m \neq 0$) independent of OD3-limit.

\section{$SL(2, Z)$ transformation}

Type IIB string theory as well as its low energy limit, the corresponding 
supergravity theory, are well-known to possess an $SL(2, Z)$ duality symmetry.
Since the (NS5, D5, D3) solution is $SL(2, Z)$ invariant, we would like to 
know what happens to the OD3-limit and OD5-limit under an $SL(2, Z)$
transformation. Under a general $SL(2, Z)$ transformation by 
the matrix given in
\eqn{8} the various fields of type IIB supergravity transform as:
\bea
g_{\mu\nu}^E &\to & g_{\mu\nu}^E, \qquad \lambda \to \frac{a\lambda + b}
{c\lambda + d}, \qquad \left(\begin{array}{c} B^{(b)}\\ 
A^{(2)}\end{array}\right) \to \left(\begin{array}{cc} d & -c\\ -b & a 
\end{array}\right)\left(\begin{array}{c} B^{(b)}\\ A^{(2)}\end{array}\right)\br
A^{(4)} &\to & A^{(4)}
\lll{30}
\eea
where $g_{\mu\nu}^E$ denotes the Einstein metric and $\lambda = \chi + 
i e^{-\phi_b}$. Under \eqn{30} the axion and the dilaton transform as
\bea
\hat{\chi} &=& \frac{(a\chi + b)(c\chi + d) + ac 
e^{-2\phi_b}}{|c\lambda +d |^2}\br
e^{\hat{\phi}_b} &=& |c\lambda + d|^2 e^{\phi_b}
\lll{31}
\eea
Since the Einstein metric $g_{\mu\nu}^E = e^{-\phi_b/2} g_{\mu\nu}$ (where
$g_{\mu\nu}$ is the string metric) remains invariant under $SL(2, Z)$ 
transformation so, the string metric would transform as
\be
d\hat{s}^2 = |c\lambda + d| ds^2
\lll{32}
\ee
If we now insist that the transformed metric should asymptotically be 
Minkowskian then the transformed metric would be given as
\be
d\hat{s}^2 = \frac{|c\lambda + d|}{\left[(c\chi_0 + d)^2 + \frac{c^2}{g_s^2}
\right]^{1/2}} ds^2
\lll{33}
\ee
We will show in the following how the metric and the dilaton would transform
under the $SL(2, Z)$ for both the OD3-limit and OD5-limit. The transformation
of the other gauge fields can be obtained in a straightforward manner.

\medskip

\underline{\it{(a) $SL(2, Z)$ transformation and OD3-limit:}} We would like 
to point out that the numerator appearing in the transformation of the
angle $\cos\psi$ in \eqn{9} may vanish for a particular choice of a set of
$SL(2, Z)$
transformations. 
It can be seen that when $c\tan\psi/g_s - (c\chi_0 + d) = 0$, 
$\cos\hat{\psi} = 0$ and $\hat{H}'' = 1$. As it is clear from the
discussion in section 3, the OD3-limit in this case would reduce to the 
NCYM-limit. So, in the following discussion we would consider the two cases
($c\tan\psi/g_s - (c\chi_0 + d) \neq 0$ and $= 0$) separately.

\medskip
\noindent\underline{(i) $c\chi_0 + d \neq c\tan\psi/g_s$:}

From the form
of $\chi$ and $e^{\phi_b}$ of OD3-limit given in \eqn{17} we find that 
\be
|c\lambda + d| = \frac{\tilde{h}''^{1/2}}{h''^{1/2}}
\lll{34}
\ee
where
\be
\tilde{h}'' = \left(\frac{c}{G_{o(3)}^2}\frac{\sqrt{1-l^2}}{l} - (c\chi_0
+d)\right)^2 + \tilde{a}^2 u^2 \left(\frac{c^2}{G_{o(3)}^4} + 
(c\chi_0 +d)^2\right)
\lll{35}
\ee
We thus find from \eqn{33} and \eqn{15} that the transformed metric has the 
form
\be
d\hat{s}^2 = \alpha' \frac{\tilde{h}''^{1/2}}{\left[(c\chi_0 +d)^2 + \frac{c^2}
{g_s^2}\right]^{1/2}}\left[-d\tilde{x}_0^2 + \sum_{i=1}^3 d\tilde{x}_i^2 +
h'^{-1}\sum_{j=4}^5 d\tilde{x}_j^2 + \frac{m}{u^2}\left(du^2 + u^2 d\Omega_3^2
\right)\right]
\lll{36}
\ee
By writing $\tilde{h}''$ as
\bea
\tilde{h}'' &=& \left(\frac{c}{G_{o(3)}^2}\frac{\sqrt{1-l^2}}{l} - (c\chi_0
+d)\right)^2 (1 + \hat{a}^2 u^2)\br
&=& \left(\frac{c}{G_{o(3)}^2}\frac{\sqrt{1-l^2}}{l} - (c\chi_0
+d)\right)^2 \hat{h}''
\lll{36a}
\eea
where
\be
\hat{a}^2 = \frac{\left[\frac{c^2}{G_{o(3)}^4} + (c\chi_0 +d)^2\right]}
{\left(\frac{c}{G_{o(3)}^2}\frac{\sqrt{1-l^2}}{l} - (c\chi_0 + d)\right)^2}
\tilde{a}^2
\lll{36b}
\ee
we can write the metric in \eqn{36} precisely in the same form as that of the
OD3-limit in \eqn{15} i.e.
\be
d\hat{s}^2 = \alpha' \hat{h}''^{1/2}
\left[-d\hat{x}_0^2 + \sum_{i=1}^3 d\hat{x}_i^2 +
h'^{-1}\sum_{j=4}^5 d\hat{x}_j^2 + \frac{\hat{m}}{u^2}\left(du^2 + 
u^2 d\Omega_3^2
\right)\right]
\lll{36c}
\ee
where
\be
\hat{m} = \frac
{\left(\frac{c}{G_{o(3)}^2}\frac{\sqrt{1-l^2}}{l} - (c\chi_0 + d)\right)}
{\left[(c\chi_0 + d)^2 + \frac{c^2}{G_{o(3)}^4}\right]^{1/2}} m
\lll{36d}
\ee
and also we have rescaled the coordinates as
\be
\hat{x}_{0,1,\ldots,5} = \frac
{\left(\frac{c}{G_{o(3)}^2}\frac{\sqrt{1-l^2}}{l} - (c\chi_0 + d)\right)^{1/2}}
{\left[(c\chi_0 + d)^2 + \frac{c^2}{G_{o(3)}^4}\right]^{1/4}} 
\tilde{x}_{0,1,\ldots,5}
\lll{36e}
\ee
Note that \eqn{36b} gives the transformation of $\tilde{a}^2 = 
\alpha'_{\rm eff}/(ml)$ under $SL(2, Z)$, however, $a^2 = 
l\alpha'_{\rm eff}/m $ remains invariant.

The form of the dilaton can be obtained from \eqn{31} and \eqn{17} as,
\be
e^{\hat{\phi}_b} = \hat{G}_{o(3)}^2 \frac{\hat{h}''}{h'^{1/2}} \frac{\hat{l}}
{\hat{a} u}
\lll{37}
\ee
Where the coupling constant of the $SL(2, Z)$ transformed OD3-theory 
is given by
\be
\hat{G}_{o(3)}^2 = \frac{1}{G_{o(3)}^2}\left[(c\chi_0 + d)^2 G_{o(3)}^4 + c^2
\right]
\lll{38}
\ee
and
\be
\hat{l} = \cos\hat{\psi} = \frac
{\left(\frac{c}{G_{o(3)}^2}\frac{\sqrt{1-l^2}}{l} - (c\chi_0 + d)\right)}
{\left[(c\chi_0 + d)^2 + \frac{c^2}{G_{o(3)}^4}\right]^{1/2}} \cos\psi
\lll{38a}
\ee
We note that the metric and the dilaton have the same forms as those of the
OD3-limit obtained in eqs.\eqn{15} and \eqn{17} in the previous section. From
\eqn{35}, it is clear that no matter whether $c\chi_0 + d \ne 0$ ($\chi_0$
is irrational) or $c\chi_0 + d = 0$ ($\chi_0$ is rational) the forms of the
metric and the dilaton in \eqn{36} and \eqn{37} are always the same as those
of the original OD3-limit. The gauge fields of the $SL(2, Z)$ transformed 
OD3-theory can also be obtained from eqs.\eqn{30}. We thus conclude that
OD3-theory is self-dual. For rational $\chi_0$, the metric and the dilaton
are given by the same expressions as in \eqn{36c} and \eqn{37} with a
simpler form of $\tilde{h}''$, $\hat{a}^2$, $\hat{m}$, 
$\hat{x}_{0,1,\ldots,5}$ and $\hat{G}_{o(3)}^2$. They are given in 
the following
\bea
\tilde{h}'' &=& \frac{c^2}{G_{o(3)}^4}\frac{1-l^2}{l^2}(1+\hat{a}^2u^2)\br
\hat{a}^2 &=& \frac{l^2}{1-l^2}\tilde{a}^2, \qquad \hat{m} = \frac{\sqrt{
1-l^2}}{l}m\br
\hat{x}_{0,1,\ldots,5} &=& \frac{(1-l^2)^{1/4}}{\sqrt{l}}
\tilde{x}_{0,1,\ldots,5},
\qquad \hat{G}_{o(3)}^2 = \frac{c^2}{G_{o(3)}^2}
\lll{39}
\eea
These are the various variables and parameters for the supergravity dual of
S-dual OD3-theory. We note that since $a^2 = l\alpha'_{\rm eff}/m$ remains
invariant and $l$, $m$ transform exactly in the same way (see \eqn{36d} and
\eqn{38a}), so, $\alpha'_{\rm eff}$ is invariant.

\medskip
\noindent\underline{(ii) $c\chi_0 + d = c\tan\psi/g_s$:}

We have already mentioned that in this case $\cos\hat{\psi} = 0$ or in other
words, $\hat{m}$, the transformed number of NS5-branes vanishes. So, in the
final theory the axion $\hat{\chi}$ and its asymptotic value $\hat{\chi}_0$
must vanish also. This indicates that $\chi_0$ in the original theory must
be rational i.e. $c\chi_0 + d = 0$. This also implies that $\sin\psi$ in the
original theory vanishes i.e. there are no D5-branes in the original theory.
From \eqn{35} we notice that when the above condition is satisfied we get
\be
\tilde{h}'' = \frac{\tilde{a}^2 u^2 c^2}{G_{o(3)}^4}
\lll{44A}
\ee
Then the metric in \eqn{32} has exactly the same form as in \eqn{26}. Also,
from \eqn{31}, the dilaton is given by,
\be
e^{\hat{\phi}_b} = \hat{g}_s \left(\frac{u}{R}\right)\frac{\tilde{b}}
{h'^{1/2}}
\lll{44B}
\ee
where $\hat{g}_s = c^2/g_s$ and $h' = 1 + a^2 u^2$, with $a^2 = 
\alpha'_{\rm eff}/(\hat{n}\hat{g}_s) = \tilde{b}/(\hat{n}\hat{g}_s),\,\, 
R = \tilde{b}/a$. Note that in this case $a^2 = \tilde{a}^2 $ since $l = 1$. 
This is
precisely the NCYM limit, we discussed in section 3, with $\tilde{b}$ as the
non-commutativity parameter and $g_{NCYM}^2 = (2\pi)^3 \hat{g}_s$, the 
Yang-Mills coupling. Thus we find NCYM-limit from a special set of $SL(2, Z)$
transformations of OD3-limit for which $\chi_0$ is rational. Therefore,
OD3-theory and NCYM theory in 5+1 dimensions can be related as strong-weak
duality symmetry as discussed in  
\cite{gopms}.
 
\medskip

\underline{\it (b) $SL(2, Z)$ transformation and OD5-limit:} Unlike in the
previous case, we note here that the expression $c\tan\psi/g_s - (c\chi_0 + d)$
can not become equal to zero while considering OD5-limit. The reason is, as
we have already mentioned, when the above expression vanishes, it implies that
$\sin\psi = 0$ in the original theory i.e. there are no D5-branes. In other
words, since $\cos\psi = 1$ in this case we can not take OD5-limit in the
original theory. So, we take $c \tan\psi/g_s - (c\chi_0 + d) \neq 0$.
Now proceeding exactly 
in the same way as in subsection $(a)$ we first find the form of 
$|c\lambda + d|$ from the expressions of $\chi$ and $e^{\phi_b}$ in \eqn{24}
as,
\be
|c\lambda + d| = \frac{\tilde{h}''^{1/2}}{h''^{1/2}}
\lll{41}
\ee
where
\be
\tilde{h}'' = \left[\frac{c}{G_{o(5)}^2} - (c\chi_0 +d)\right]^2 +
\tilde{a}^2 u^2 (c\chi_0 + d)^2
\lll{42}
\ee
Note that since the form of $\tilde{h}''$ changes completely for irrational
$\chi_0$ and rational $\chi_0$, we study these two cases separately.

\noindent\underline{\it Irrational $\chi_0$}

When $c\chi_0 + d$ is not equal to zero, we find from \eqn{33} and \eqn{22}
that the transformed metric takes the form:
\be
d\hat{s}^2 = \alpha' \frac{\tilde{h}''^{1/2}}
{(c\chi_0 + d)}\left[-d\tilde{x}_0^2
+ \sum_{i=1}^5 d\tilde{x}_i^2 + \frac{m}{u^2}\left(du^2 + u^2 d
\Omega_3^2\right)\right]
\lll{43}
\ee
Redefining
\bea
\tilde{h}'' &=& \left[\frac{c}{G_{o(5)}^2} - (c\chi_0 + d)\right]^2 
(1+\hat{a}^2
u^2)\br
&=& \left[\frac{c}{G_{o(5)}^2} - (c\chi_0 + d)\right]^2 \hat{h}''
\lll{44}
\eea
where
$\hat{h}'' = 1 + \hat{a}^2 u^2$ and
\be
\hat{a}^2 = \frac{(c\chi_0 + d)^2}{\left[\frac{c}{G_{o(5)}^2} - (c\chi_0 + d)
\right]^2} \tilde{a}^2
\lll{45}
\ee
we can rewrite the metric in \eqn{43} exactly in the same form as that of 
OD5-limit, namely,
\be
d\hat{s}^2 = \alpha' \hat{h}''^{1/2}\left[-d\hat{x}_0^2
+ \sum_{i=1}^5 d\hat{x}_i^2 + \frac{\hat{m}}{u^2}\left(du^2 + u^2 d
\Omega_3^2\right)\right]
\lll{46}
\ee
where
\be
\hat{m} = \frac{\left[\frac{c}{G_{o(5)}^2} - (c\chi_0 + d)
\right]}{(c\chi_0 + d)} m
\lll{47}
\ee
we have also redefined the coordinates as,
\be
\hat{x}_{0,1,\ldots,5} = \frac{\left[\frac{c}{G_{o(5)}^2} - (c\chi_0 + d)
\right]^{1/2}}{(c\chi_0 + d)^{1/2}} \tilde{x}_{0,1,\ldots,5}
\lll{48}
\ee
The transformed dilaton can be obtained from \eqn{31} and \eqn{24} as,
\be
e^{\hat{\phi}_b} = \hat{G}_{o(5)}^2 \frac{\hat{h}''}{\hat{a}u}
\lll{49}
\ee
where
\be
\hat{G}_{o(5)}^2 = (c\chi_0 + d) \left[\frac{c}{G_{o(5)}^2} - 
(c\chi_0 + d)\right] G_{o(5)}^2
\lll{50}
\ee
Here we note that since $\cos\varphi = \tilde{l} = a/\tilde{a} =$ invariant
under $SL(2, Z)$, so both $\tilde{a}$ and $a$ must transform in the same way.
The transformation of $\tilde{a}$ is given in \eqn{45}. However, $\cos\psi
= \epsilon = m/(\sqrt{m^2 + (n + \chi_0 m)^2 g_s^2})$ is not $SL(2, Z)$ 
invariant, but it must transform as $m$ i.e.
\be
\cos\hat{\psi} = \hat{\epsilon}= \frac{\left[\frac{c}{G_{o(5)}^2} - 
(c\chi_0 + d)
\right]}{(c\chi_0 + d)} \cos\psi
\lll{51}
\ee
Thus we observe from \eqn{46} and \eqn{49} that the transformed metric and the
dilaton have exactly the same form as those of OD5-limit obtained in \eqn{22}
and \eqn{24}. Also since $\tilde{a}^2 = \alpha'_{\rm eff}/(m \tilde{l})$, so
the effective tension of the $SL(2, Z)$ transformed OD5-theory would be
given as,
\be
\hat{\alpha'}_{\rm eff} = \frac{(c\chi_0 + d)}{\left[\frac{c}{G_{o(5)}^2} - 
(c\chi_0 + d)
\right]} \alpha'_{\rm eff}
\lll{52}
\ee
The forms of the other gauge fields of the $SL(2, Z)$ dual OD5-theory can be
obtained from \eqn{30}.

\medskip

\noindent\underline{\it Rational $\chi_0$}

Now if $\chi_0$ is rational i.e. $c\chi_0 + d = 0$, then from \eqn{41} we have
\be
|c\lambda + d| = \frac{\tilde{h}''^{1/2}}{h''^{1/2}}
\lll{53}
\ee
where $\tilde{h}'' = c^2/G_{o(5)}^4$. So, the metric and the dilaton would
now take the forms,
\bea
d\hat{s}^2 &=& \left[-d\tilde{x}_0^2
+ \sum_{i=1}^5 d\tilde{x}_i^2 + \alpha'_{\rm eff}\frac{m}{u^2}
\left(du^2 + u^2 d
\omega_3^2\right)\right]\br
e^{\hat{\phi}_b} &=& \frac{c^2}{G_{o(5)}^2} \frac{1}{\tilde{a}u}
\lll{54}
\eea
Here $G_{o(5)}^2$ is just a finite quantity which can be absorbed in 
$\tilde{a}$. We have also redefined the coordinates $\tilde{x}_{0,1,\ldots,
5}$ by $\sqrt{\alpha'_{\rm eff}}\tilde{x}_{0,1,\ldots,5}$ and 
$\alpha'_{\rm eff} = \alpha'/\epsilon =$ finite. Note also that although
$e^{\hat{\phi}_b}$ is finite, the coupling constant for the dual theory 
$\hat{G}_{o(5)}^2 = \hat{g}_s = (c^2/G_{o(5)}^2)\epsilon \to 0$. Eq.\eqn{54}
represents precisely the supergravity dual of little string theory
\cite{imsy,abks}. We thus
conclude that the S-dual of OD5-theory is the little string theory.

\section{Conclusion}

To summarize, we have studied in this paper the various decoupling limits
of an $SL(2, Z)$ invariant bound state of the type (NS5, D5, D3) in type
IIB supergravity. This solution can also be regarded as NS5-branes in the
presence of both a 4-form and a 6-form RR electric gauge fields. In particular,
we have identified an OD3-limit and an OD5-limit for this solution. In these
decoupling limits (NS5, D5, D3) solution represent the supergravity dual of
OD3-theory and OD5-theory respectively. In both the cases we obtained
noncommutative theories and are different from the usual OD3 and OD5 theories. 
We have mentioned that when NS5-branes
are absent, the OD3-limit reduces to an NCYM limit. But we do not find an
independent NCYM limit in the presence of NS5-branes. We then studied the
$SL(2, Z)$ transformation of both OD3-limit and OD5-limit. The generic 
$SL(2, Z)$
transformation of OD3-limit always gives another OD3-limit with different
set of parameters irrespective of whether the asymptotic value of axion
is irrational or not. When the asymptotic value of the axion is rational the
two OD3-theories are related to each other by strong-weak duality symmetry.
We thus conclude that OD3-theory is self-dual. However, under a special set
of $SL(2, Z)$ transformations we find that the OD3-limit reduces to NCYM-limit.
But for these set of $SL(2, Z)$ transformations the transformed charge of
NS5-brane vanishes. Thus it is not surprising that we get an NCYM-limit for
these transformations as we have already mentioned in section 3. In this
case $\chi_0$ is rational and so, the OD3-theory and NCYM-theory are 
related by strong-weak duality symmetry  
and this case has already been studied in \cite{gopms}. On the other hand, for
OD5-limit, we find that when the asymptotic value of axion is irrational
a generic $SL(2, Z)$ transformation gives another OD5-limit with different
set of parameters characterizing the $SL(2, Z)$ transformed OD5-theory. But
when $\chi_0$ is rational OD5-limit reduces to little string theory limit.
So, we conclude that OD5-theory and little string theory are related to each
other by type IIB S-duality symmetry. 

\section*{Acknowledgements}

We would like to thank Jianxin Lu for very useful comments and a critical
reading of the manuscript.

\vfil
\eject

\end{document}